\documentclass[aps,pra,showpacs,twoside,twocolumn,10pt]{revtex4-1}
\usepackage[colorlinks=true, citecolor=red, urlcolor=blue ]{hyperref}
\usepackage{epsfig,newlfont,amssymb,amsfonts,amsmath,bm,subfigure,palatino,mathtools,amsthm,braket,times,soul}
\usepackage{multirow}

\begin{document}


\title{Rating the performance of noisy teleportation using fluctuations in fidelity}

\author{Saptarshi Roy\(^1\) and Arkaprabha Ghosal\(^{2}\)}

\affiliation{\(^1\)Harish-Chandra Research Institute, HBNI, Chhatnag Road, Jhunsi, Allahabad 211 019, India}
\affiliation{\(^2\)Department  of  Physics  and  Center  for  Astroparticle  Physics  and  Space  Science,Bose  Institute,  Bidhan  Nagar  Kolkata  -  700091,  India}

\begin{abstract}
Quantum teleportation is one of the most pioneering features of the quantum world. 
Typically, the quality of a teleportation protocol is solely judged by its average fidelity. In this work, we analyze the performance of teleportation in terms of both fidelity and the deviation in fidelity. Specifically, we define a quantity called teleportability score, which incorporates contributions from both the fidelity and its deviation. It also takes into account the sensitivity one requires for a protocol in which the teleportation of a quantum state is required in one or many intermediate steps.  
We compute the teleportability score in the noiseless scenario and find that it increases monotonically with the entanglement content of the considered pure resource states. The result remains same even if we consider an n-chain repeater-like configuration. However, in the presence of noise, the teleportability score, can sometime display a nonmonotonic behaviour with respect to the entanglement content of the initially shared resource state. Specifically, under local bit-flip and bit-phase-flip noise, lesser entangled states can have higher teleportability score for certain choice of system parameters.  In the presence of global depolarizing noise, for low entangled resource states and high sensitivity requirements, the noisy states can have better a teleportability score in comparison to the noiseless scenario.

\end{abstract}

\maketitle

\section{Introduction}

Quantum information science is continually revolutionizing the fields of communication and computation. Pioneering quantum protocols include quantum cryptography \cite{BB84,Crypto}, quantum dense coding \cite{DC}, quantum error correction \cite{error_correction}, quantum teleportation \cite{bennett1993} etc. The field of communication is one of the major beneficiaries due to the advent of these protocols, which have also been implemented experimentally in a variety of physical systems \cite{communication_exp}. For example, quantum communication protocols like dense coding \cite{DC} out perform the
 classical ones (in terms of capacity) by a factor of $2$. Furthermore, for secure communication, the security of quantum protocols \cite{BB84,Crypto}  is guaranteed by the laws of physics, unlike the classical case where security is derived from exponential complexity of some mathematical problem. These examples establish superiority of the protocols in the quantum domain.

Among the quantum communication protocols, quantum teleportation \cite{bennett1993,tele_review} has been one of the central interests of study. It has been studied quite extensively both theoretically \cite{murao1998} and experimentally   \cite{tele_photon_exp,tele_photon_exp2,tele_contvar_exp,tele_ion_exp,tele_exp_coldatom,tele_nmr_exp,tele_supercond}. However, almost all of these works, characterize the performance of a given teleportation protocol solely by the average fidelity it yields. But such characterization is very limited since it does not incorporate the effect of fluctuations. Different input states can have widely varying fidelities keeping the average value of fidelity fixed. Such high fluctuations can be detrimental during implementation of some quantum information processing protocols, like that of quantum gates in quantum computation \cite{ref11,ref12}.  Therefore, not only the average fidelity but also the distribution of the fidelity for various input states is what that determines the performance of teleportation.
Analysis of fluctuation and distribution of fidelities have been studied \cite{gatefid1} in the context of the performance of single qubit gates. Similar investigations for quantum gates have also been carried out using higher order moments \cite{gatefid2}. In the avenue of quantum teleportation, studies of fluctuations using deviation in fidelities was first introduced in \cite{dev1}, and was later formalized in \cite{fid_dev}, where they analyzed quality of teleportation in the plane of fidelity and its deviation (see also \cite{g1,g2}).

In this work, we seek a quantitative answer while comparing the performance of two resource states (noiseless or noisy) for a given teleportation protocol. For this purpose, we have introduced a new performance indicator of quantum teleportation, `teleportability score', which, apart from the average fidelity, also incorporates its fluctuations while rating. It also takes into account the sensitivity requirements of a particular setup while rating. This is so because not all experimental setups are equally sensitive to fluctuations in fidelity.  Depending on the sensitivity requirements, one adjusts the relative weights of fidelity and its deviation in the teleportability score. We analyze the teleportability score of resource states subjected to local and global noises and contrast it with the clean (noiseless) case for different values of sensitivity requirements.

Firstly, in the noiseless scenario, we find that the fidelity deviation decreases with increasing values  of fidelity, thereby the teleportability score is always higher for more entangled states. However, in the presence of local noise, such ordering of states with respect to entanglement is not always present. For certain parameter ranges, we find that lesser entangled states can yield better or equally good teleportability scores in comparison
 to states having higher entanglement content. This is due to the fact that, unlike in the noiseless scenario, in presence of noise, both fidelity as well as its fluctuation might increase when higher entangled states are used for teleportation. In this paper, we investigate the teleportability score in presence of both locally and globally noisy channels.  The local noise models \cite{preskil,Nielsen_chuang} considered in this 
 manuscript are namely, bit-flip, phase-flip, bit phase-flip, amplitude damping and depolarizing noise. We consider these noises to act locally on both the qubits of the shared entangled state. We also compute the teleportability score in presence of global depolarizing (white) noise and when the resource state suffers from both global and local noises.

The paper is organized as follows. We formally define the teleportability score in Sec. \ref{sec:tele_score}
The analysis for the noiseless scenario is done in Sec. \ref{sec:noiseless}. The fidelity and its deviation for a $n$-chain repeater-like setting is computed in Sec. \ref{sec:n_chain}. The investigation of the teleportability score in the presence of noise is presented in Sec. \ref{sec:noise}. The issue with local noise is dealt in Sec. \ref{sec:local_noise}. The case studies with bit-flip, phase-flip, and bit phase-flip noises are given in Sec. \ref{sec:bitflip} - Sec.\ref{sec:bitphaseflip}.
The issue of global noise and combination of local and global noises are addressed in Sec. \ref{sec:global}.  Finally we conclude in Sec. \ref{sec:conclusion}.

\section{Teleportability score}
\label{sec:tele_score}

The average fidelity of a given teleportation scheme is given by
\begin{eqnarray}
F = \int \text{d}\phi \langle \phi | \rho_{\phi} | \phi \rangle,
\end{eqnarray}
where $|\phi\rangle$ is the arbitrary state to be teleported and $\rho_{\phi}$ is the teleported state with $|\phi\rangle$ at the input. The averaging is performed over all possible input states. Whenever we mention fidelity, $F$ in this manuscript, we refer to the average fidelity. The fidelity obtained for a particular input state $|\phi\rangle$ is simply given by $F^\phi = \langle \phi | \rho_{\phi} | \phi \rangle$. The corresponding deviation in fidelity (standard deviation) reads
\begin{eqnarray}
D &=& \sqrt{\int \text{d}\phi \langle \phi | \rho_{\phi} | \phi \rangle^2 - \Big( \int \text{d}\phi \langle \phi | \rho_{\phi} | \phi \rangle \Big)^2}, \nonumber \\
 &=& \sqrt{\int \text{d}\phi \langle \phi | \rho_{\phi} | \phi \rangle^2 - F^2}.
\end{eqnarray}

In the usual fidelity based rating of a teleportation scheme, for a given protocol, two resource states with same $F$ but different $D$'s are deemed to be equivalent. But one can easily argue that the state having a lower value of $D$ is clearly a better choice. Therefore one might expect a payoff in the rating scheme of teleportation protocols due to $D$. With the motivation of quantifying the quality of  teleportation in a more general context, we define a new quantity, teleportability score  as
\begin{eqnarray}
\tau_k = F-k.D,
\label{eq:tele_score}
\end{eqnarray}
where $F$ and $D$ are the fidelity and its deviation  and $k$ has to be chosen according to the sensitivity requirements to fluctuations in fidelity of a particular protocol in which the teleportation is used as an intermediate step. Greater sensitivity requirements would simply imply a higher pay off in the score due to $D$ which is ensured by choosing a higher value of $k$. However, practically, we cannot choose an arbitrarily large $k$. 
This simply indicates that there is a physical cutoff to the maximum amount of sensitivity we can demand out of the teleportation process. In particular, we call a $k$ value to be large (i.e. the score is highly sensitive to fluctuations) when the payoff function $k.D$ is comparable to the average fidelity $F$. This defines a scale for the sensitivity values, $k^*$, defined via the following relation:
\begin{eqnarray}
k^*.D \approx F \implies k^* \approx F/D.
\label{eq:k*}
\end{eqnarray} 
Physically it does not seem natural to consider the pay off function higher than the average fidelity. So  from a practical point of view, we can approximately provide an upper bound of $k$ as $k^*$. Therefore approximately, $k \in (0,k^*)$. We would examine the $k^*$ values both in the noiseless and noisy scenarios in subsequent sections.
Nevertheless, apart from its interpretation as the sensitivity parameter, there can be situations where $k$ assumes a mathematical significance. Suppose, we want to design a protocol which maximizes the fidelity for a fixed value of standard deviation. In such a constrained optimization problem, $k$ takes the role of a Lagrange's multiplier. Naturally, $k$ is constrained to take non negative values.

 The best fidelity obtained in a classical (entanglement \cite{horodecki2009} free) scheme is $F^{cl} = \frac{2}{3}$ \cite{classical1,classical2}.
We would carry out our analysis of rating teleportation performance using teleportability score only for those states those yield a nonclassical average fidelity of teleportation, i.e.,  $F > F^{cl}$. Naturally, if $F \leq F^{cl}$, its anyway not deemed to be useful for quantum teleportation.
Note that  following the same classical fidelity maximizing protocol,  the corresponding fidelity deviation is also easily calculated to be $D^{cl} = \frac{1}{3\sqrt{5}}$. So, we define the \emph{classical value} of the teleporatability score for a given $k$, following Eq. \eqref{eq:tele_score}, as
\begin{eqnarray}
\tau_k^{cl} = F^{cl} - k.D^{cl} = \frac{1}{3}\big ( 2 - k/\sqrt{5} \big)
\label{eq:tele_score_classical}
\end{eqnarray}   
A state with $F > F^{cl}$ would therefore be considered to possess any quantum advantage only when for the given protocol, it's teleportability score is higher than the classical limit, i.e., $\tau_k > \tau_k^{cl}$. 
Therefore, in summary, a quantum state would be deemed to be useful for quantum teleportation iff it satisfies both $F > F^{cl}$ and $\tau_k > \tau_k^{cl}$.
Once again, the classical teleportability score is computed by considering the fidelity maximizing protocol and not via the overall maximization of the teleportability score with respect to all classical protocols. Similar strategy (with respect to the choice of the protocol) is adopted in the quantum case as well which we discuss in the subsequent sections. 

\section{The noiseless scenario}
\label{sec:noiseless}
\begin{figure}[ht]
\includegraphics[width=\linewidth]{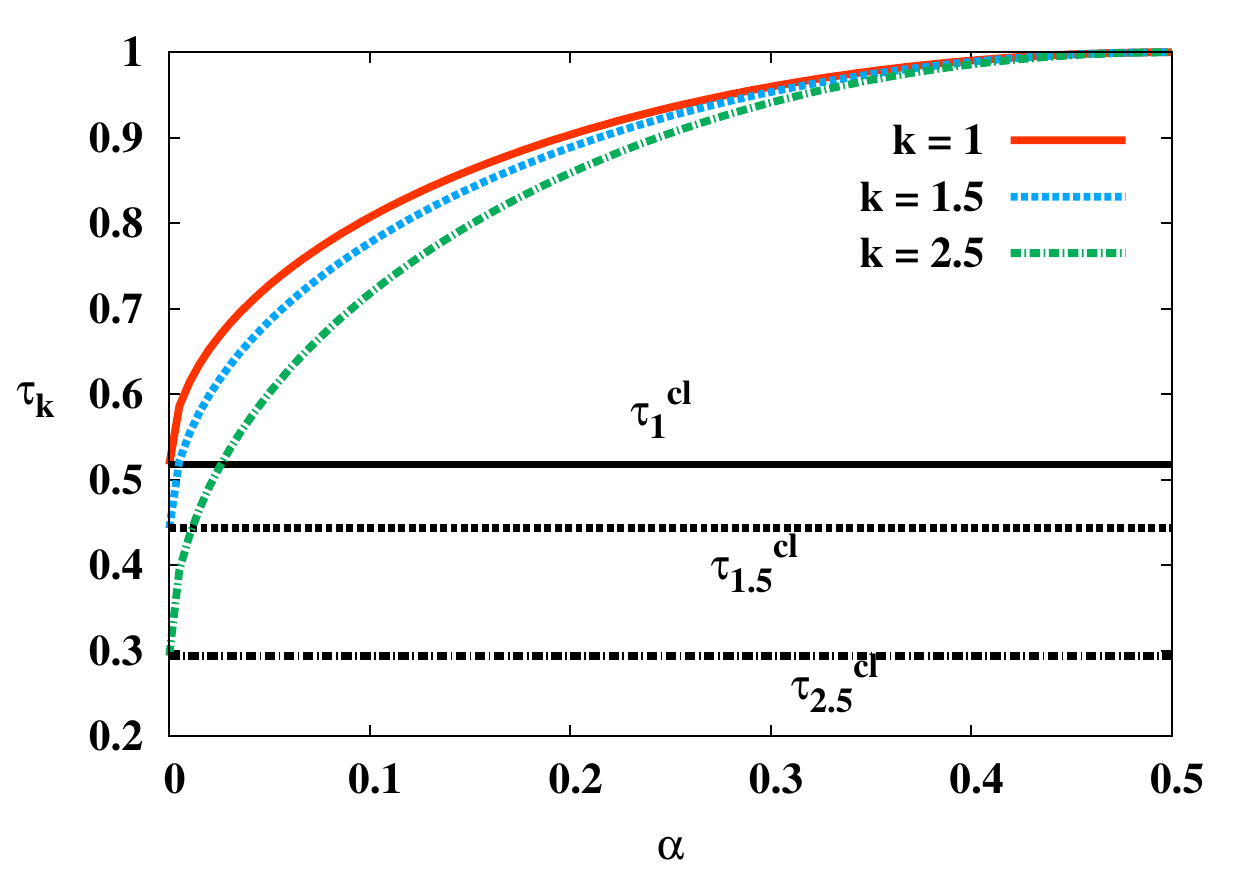}
\caption{Teleportability score for different values of $k$ in the noiseless scenario using a single shared resource state. All axes are dimensionless.}
\label{fig:noiseless}
\end{figure}

The initial shared state to be utilized for teleportation, in the Scmidt form \cite{Schmidt_decomposition}, is 
\begin{eqnarray}
|\psi^{\alpha}\rangle = \sqrt{\alpha}|00\rangle + \sqrt{1-\alpha}|11\rangle.
\label{eq:psi_alpha}
\end{eqnarray}
 The maximal singlet fraction, $f_{max}$, for this shared state is given by $1/2 + \sqrt{\alpha(1-\alpha)}$, and following the prescription given in \cite{horo_fid}, the maximal teleportation fidelity reads
\begin{eqnarray}
F_{max} = \frac{1}{3}(2f_{max}+1) = \frac{2}{3}+\frac{2}{3}\sqrt{\alpha(1-\alpha)}.
\label{eq:fid_noiseless} 
\end{eqnarray}
For brevity, we would henceforward refer $F_{max}$ as $F$ and the corresponding deviation as $D$. 
The above fidelity can be achived by performing Bell measurements \cite{Bell-basis} at the Alice's end and appropriate Pauli untaries at Bob's end after $2$-bits of classical communication, i.e., the usual teleportation protocol \cite{bennett1993}.
Note that the maximal fidelity in the usual protocol is obtained by optimizing over local unitary operations.
Specifically,  when an arbitrary state, $|\eta\rangle =$ $\cos \frac{\theta}{2}|0\rangle +e^{i\phi} \sin \frac{\theta}{2}|1\rangle$ is teleported using the above protocol, the fidelity reads
\begin{eqnarray}
F^\eta = 1 - \frac{1}{2}(1 - 2 \sqrt{\alpha(1-\alpha)})\sin^2 \theta,
\end{eqnarray}
which when uniformly averaged over the Bloch sphere parameters yields the same fidelity as in Eq. \eqref{eq:fid_noiseless}, thereby establishing the usual teleportation protocol as an optimal one (in terms of fidelity) for states of the structure as given in Eq. \eqref{eq:psi_alpha}.
   The fidelity deviation, $D$,  corresponding to the above protocol is given by
\begin{eqnarray}
D &=& \frac{1}{3\sqrt{5}}\sqrt{1 + 4 (1 - \alpha) \alpha - 4 \sqrt{(1 - \alpha) \alpha}} \nonumber \\
 	&=& \frac{1}{\sqrt{5}}(1-F).
 	\label{eq:dev_noiseless}
\end{eqnarray}
Therefore, in the noiseless scenario, the deviation in fidelity can be completely specified in terms of the average fidelity, $F$. We notice a sort of \emph{win-win} situation since in the noiseless case, greater the fidelity, lesser is the fluctuation (deviation). For the maximally entangled resource state ($\alpha = 1/2$), we get a unit teleportation fidelity, $F=1$, and the corresponding deviation falls to zero ($D=0$). In Fig. \ref{fig:noiseless}, we plot the teleportability scores for various $k$-values in the noiseless scenario and also depict the classical values of teleportability scores ($\tau_k^c$) for those $k$-values.

Therefore, the telepotibility score in the noiseless scenario reads
\begin{eqnarray}
\tau_k = F - k.D = (1+\frac{k}{\sqrt{5}})F - \frac{k}{\sqrt{5}}.
\end{eqnarray}
Note that, in the noiseless scenario, $\tau_k$ is an increasing function of $\alpha$ for all values of $k$.
 Therefore we can conclude that \emph{more entangled resource states yield higher teleportability score in the noiseless scenario.} Although the observed feature is $k$ independent, we explore the $k^*$ values in this noiseless case. If we follow Eq. \eqref{eq:k*}, one question still remain. What should be the choice of $\alpha$ while computing $k^*$ via Eq. \eqref{eq:k*}. We resolve this by selecting the $\alpha$ value that yield the lowest $k^*$ value. This generates the most conservative upper bound for $k$. Therefore, mathematically, we define 
 \begin{eqnarray}
 k^* = \min_{\alpha} \frac{F}{D},
 \end{eqnarray}
 Note that $\alpha = 0$ minimizes the $F/D$ in the noiseless scenario, and $k^*$ turns out to be $2\sqrt{5} \approx 4.5$.
 
Nevertheless, we want to mention that the result of $k$ independence holds true
  when instead of one shared state, we have a repater-like configuration consisting of $n$ entangled states.  We shall investigate how this situation changes in presence of noise in the succeeding sections. But before that, we would investigate the fidelity and deviation in a repeater like scenario.

\begin{figure}[h]
\includegraphics[width=\linewidth]{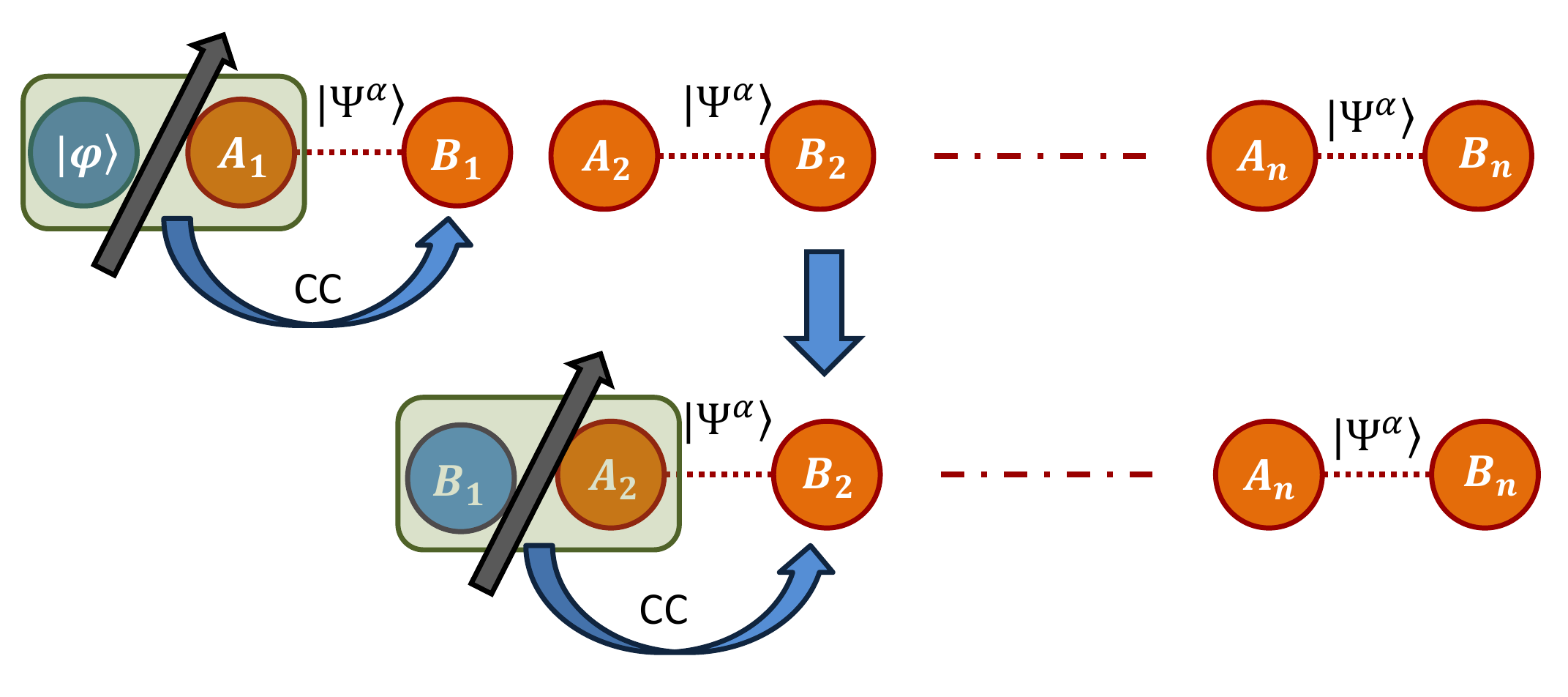}
\caption{The first two steps of teleportation with $n$-entangled states in a repeater-like setting. Each element of the chain consists of the same state $|\psi^\alpha\rangle = \sqrt{\alpha}|00\rangle + \sqrt{1-\alpha}|11\rangle$. An arbitrary state, $|\phi\rangle = a|0\rangle + b|1\rangle$ is teleported via the chain by successive application of the teleportation protocol ($n$-steps). Note, the post measurement state after implementing the protocol for $m$ states in the chain becomes the input state for the $m+1^{\text{th}}$ state. The ``CC" in the figure denotes the classical communication of the clicking results during the Bell measurement.}
\label{fig:nparty}
\end{figure}
 \subsection{The n-chain configuration}
 \label{sec:n_chain}
In a repeater like setting \cite{q_repeater}, the teleportation protocol is executed in succession for each entangled state of the chain. The teleported state after the $m^{\text{th}}$ states becomes the input state for the $m+1^{\text{th}}$ state of the chain. See Fig. \ref{fig:nparty}. The fidelity  for the $n$-chain configuration whereby each segment is constructed of the state $|\psi^\alpha\rangle = \sqrt{\alpha}|00\rangle + \sqrt{1-\alpha}|11\rangle$, when an arbitrary state, $|\eta\rangle =$ $\cos \frac{\theta}{2}|0\rangle +e^{i\phi} \sin \frac{\theta}{2}|1\rangle$ is to be teleported through the chain, is given by
\begin{eqnarray}
F_n^\eta  &=& 2^{n+1}   \alpha^{n/2}(1-\alpha)^{n/2} \cos^2 \frac{\theta}{2}\sin^2 \frac{\theta}{2} \nonumber \\
 &+& \sum\limits_{k=0}^{n} \binom{n}{k}\alpha^k (1-\alpha)^{n-k} \big( \cos^4 \frac{\theta}{2} + \sin^4 \frac{\theta}{2} \big) \nonumber \\
 &=&  (1 - \frac{\sin^2 \theta}{2}) + 2^{n-1}   \alpha^{n/2}(1-\alpha)^{n/2} \sin^2 \theta.
\label{eq:Nlinearch}
\end{eqnarray}
The corresponding average fidelity and deviation reads
\begin{eqnarray}
F_n &=& \frac{2}{3} + \frac{2^n}{3}\lbrace\alpha(1-\alpha)\rbrace^{\frac{n}{2}} \nonumber \\
D_n &=& \frac{1}{3\sqrt{5}}\sqrt{1 + 2^{2n}\lbrace\alpha(1-\alpha)\rbrace^n + 2^{n+1}\lbrace\alpha(1-\alpha)\rbrace^{\frac{n}{2}}} \nonumber \\
 &=& \frac{1}{\sqrt{5}}(1-F_n).
 \label{eq:nchain}
\end{eqnarray}
Note that the expressions in Eq. \eqref{eq:nchain} reduces to that given in Eqs. \eqref{eq:fid_noiseless} and \eqref{eq:dev_noiseless} on substituting $n = 1$. Thus in the noiseless scenario, even for the $n$-chain setting, we have qualitatively the same results as in the case of a single shared entangled state. In the rest of the manuscript where we deal with noisy channels, we would restrict our investigation of the teleportability score for teleportation with a single noisy channel.

\section{The noisy scenario}
\label{sec:noise}

The presence of noise is ubiquitous in nature and inevitably affects the performance of any protocol quantum or classical. The noiseless situation is rather ideal  and one must go beyond the noiseless assumption to make predictions in a more realistic setting. In the context of this manuscript, the omnipresence of noise demands the re-examination of the performance of teleportation  in the presence of imperfections. In this attempt, we consider noisy resource states and compute their teleportability score for various ranges of noise parameters. We consider the situation of both local and global noises and investigate their impact on the quality of teleportation. Situations where both global and local noises act together are also analyzed in subsequent sections.
\begin{figure}[h]
\includegraphics[width=\linewidth]{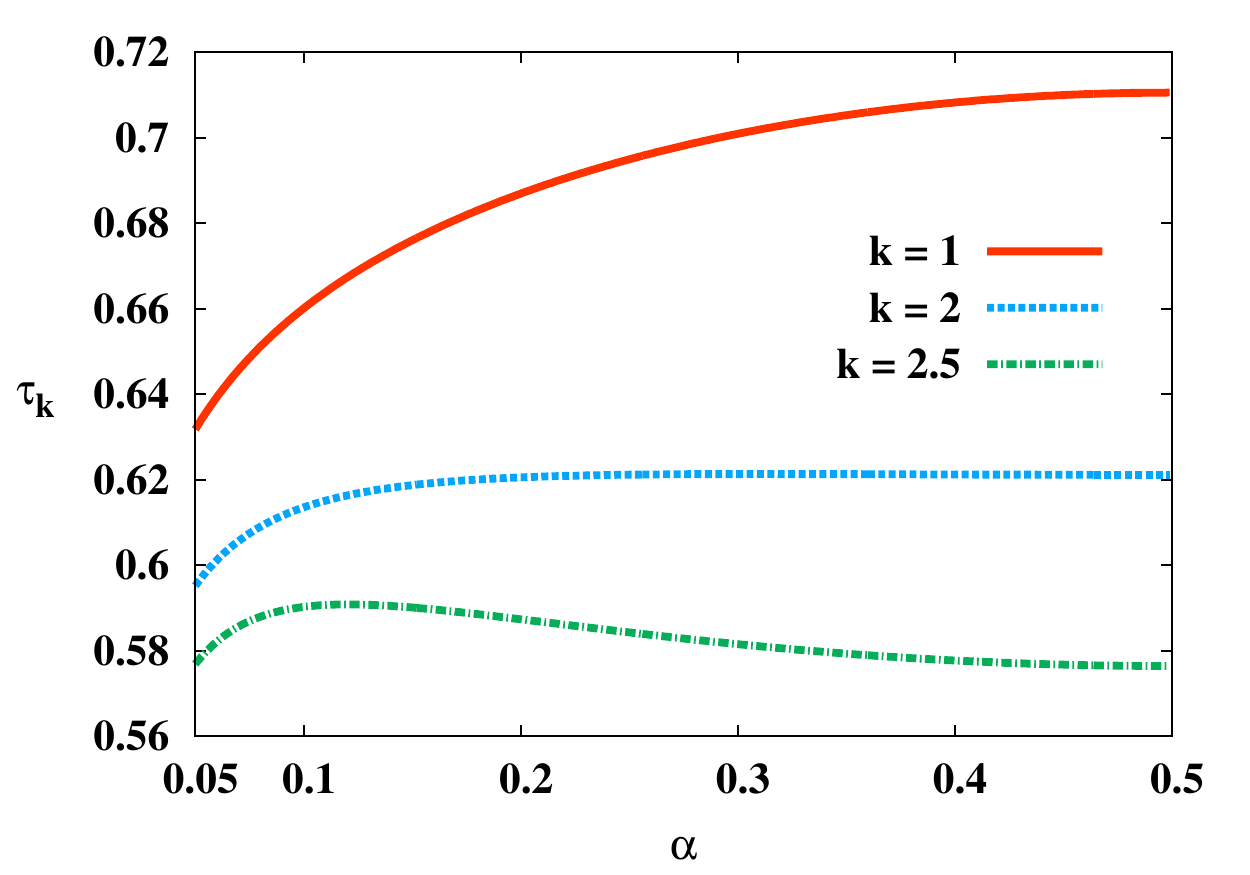}
\caption{Teleportability score for different values of $k$ for local bit-flip and bitphase-flip noises with $p=0.7$ and $q=1$. Although the average fidelity increases with $\alpha$, $\tau_k$, depending on how high $k$ is might decrease with increasing $\alpha$. The offset of $0.05$ in $\alpha$ assures that the noisy resource yields nonclassical teleportation. All the axes are dimensionless.}
\label{fig:bitflip}
\end{figure}

\subsection{Local noise}
\label{sec:local_noise}
In this section, we consider the situation where the resource state for teleportation suffers from local noise in both its qubits with different rates. The local noise models we have considered are namely bit flip, phase flip, bit-phase flip, amplitude damping channel, and phase damping channel \cite{preskil}. The a comparative study of the response of teleportability score to these different kinds of local noise is presented. Our analysis reveal a counter intuitive feature in the presence of local bit flip noise where for certain range of noise parameter and sensitivity, we get \emph{more teleportability score with a lesser entangled resource state.} In the noisy scenario, the \emph{win-win} situation of higher fidelity with lesser deviation is lost and the teleportability score becomes a much more realistic quantifier of the performance of 
teleportation. Therefore, teleportability score, which assumed a passive role as a quantifier of the `goodness' of teleportation in the noiseless scenario, attains an active role and   becomes much more physically relevant in presence of noise.

\subsubsection{Bit-flip noise}
\label{sec:bitflip}
The bit-flip noise can be modeled by the action of the Pauli operator $\sigma^x$. 
Unsurprisingly, it flips the state $\ket{0}$ to $\ket{1}$ and vice-versa. Given an arbitrary state, $\rho$, it keeps it unaltered with a probability $p$, and flips it's bits
with a probability of $1 - p$. Naturally, $0 \leq p \leq 1$. Note that the Krauss operators for the noise model can be written as
\begin{eqnarray}
K_0^{\text{bf}} &=& \sqrt{pq} ~\mathbb{I}\otimes \mathbb{I}, ~K_1^{\text{bf}} = \sqrt{p(1-q)} ~\mathbb{I} \otimes \sigma^x, \nonumber \\
K_2^{\text{bf}} &=& \sqrt{(1-p)q} ~\sigma^x  \otimes \mathbb{I}, ~K_3^{\text{bf}} = \sqrt{(1-p)(1-q)} ~\sigma^x \otimes \sigma^x, \nonumber \\
\end{eqnarray}
where the superscript bf denotes bit-flip noise, $\sigma_{1(2)}^x = \begin{bmatrix}
0 & 1 \\
1 & 0
\end{bmatrix}$, and $\mathbb{I}=\begin{bmatrix}
1 & 0 \\
0 & 1
\end{bmatrix}$. The corresponding state evolves as
\begin{eqnarray}
\rho \xrightarrow{\text{bit flip}} \sum_{i=0}^3 K_i^{\text{bf}} \rho (K_i^\text{bf})^\dagger.
\end{eqnarray}

Therefore in the presence of bit flip noise in both the qubits with probabilities $p$ and $q$, an arbitrary resource state $\rho_{12}$ becomes 
\begin{eqnarray}
\rho_{12} \xrightarrow{\text{bit flip}} pq \rho_{12} &+& (1-p)q \sigma^x_1 \rho_{12} \sigma_1^x + p(1-q) \sigma^x_2 \rho_{12} \sigma_2^x \nonumber \\ &+& (1-p)(1-q) \sigma^x_1 \sigma^x_2 \rho_{12} \sigma^x_2 \sigma_1^x,
\end{eqnarray} 
where the subscript $l$ in $\sigma^{\hat{n}}_l$ denotes on which qubit of $\rho_{12}$ it acts on.
When an arbitrary state, $|\eta\rangle =$ $\cos \frac{\theta}{2}|0\rangle +e^{i\phi} \sin \frac{\theta}{2}|1\rangle$, is to be teleported through such a channel, the fidelity of the output state with Bob with $|\eta\rangle$ reads
\begin{eqnarray}
F^\eta &=& \big(pq +(1-p)(1-q)\big) \big[ 1 - \frac{1}{2}(1 - 2 \sqrt{\alpha(1-\alpha)})\sin^2 \theta\big] \nonumber \\
&+& \frac{1}{2}(p + q - 2pq)\big[1 + 2\cos 2\phi \sqrt{\alpha(1-\alpha)}\big]\sin^2 \theta.
\label{eq:fid_bitflip}
\end{eqnarray}
 Although the average fidelity and fidelity deviation can be computed easily from the above equation, but the corresponding expressions, especially for the deviation, becomes cumbersome. So, to simplify matters we assume $q=1$. Note that we would have obtained the same results if we have instead assumed $p=1$.  The computed fidelity reads   
 \begin{eqnarray}
 F^{\text{bit-flip}} = \frac{1}{3} \big( 1-p + 2p \big(1 + \sqrt{(1-\alpha ) \alpha}\big) 
 \end{eqnarray}
 and the corresponding deviation is given by
\begin{widetext}
\small
\begin{eqnarray}
D^{\text{bit-flip}} = \frac{1}{3\sqrt{5}}\sqrt{1+4 (1-\alpha ) \alpha -4 \sqrt{(1-\alpha ) \alpha }+4 (1-p)^2 (1+4 (1-\alpha ) \alpha - 2 \sqrt{(1-\alpha ) \alpha}) - 4 (1-p) (1+2 (1-\alpha ) \alpha -3 \sqrt{(1-\alpha ) \alpha })} \nonumber \\
\label{eq:bitflip_F,D}
\end{eqnarray}
\normalsize
\end{widetext}
Clearly, the fidelity, $F^{\text{bit-flip}}$, remains an increasing function of $\alpha$ ($0 \leq \alpha \leq 1/2$). However, unlike the noiseless case, $D^{\text{bit-flip}}$, after some initial non-monotonicities, also increases with increasing $\alpha$. Therefore for high sensitivity requirements, the weight of the payoff term outweighs the gain in fidelity with increasing $\alpha$. So, when the resource states suffer from local bit-flip noise,  there are situations in which the hierarchy of the teleportability score in terms of entanglement is lost (See Fig. \ref{fig:bitflip}). Thus, in the presence of local bit flip noise, for certain choice system parameters, we get \emph{more teleportability score with less entanglement.}

Note that for the choice of noise parameters chosen for Fig. \ref{fig:bitflip}, i.e., $p = 0.7$, and $q = 1$, we get $k^* \approx 9$. The $k^*$ value is almost double of what was computed in the noiseless case. So the choice of $k = 2.5$ is much lower than the large $k = k^*$ bound. Moreover, in contrast to the noiseless scenario, the $k^*$ obtained in this case is for $\alpha = 0.5$. This is due to the fact that unlike the noiseless case, in presence of bit-flip noise, both fidelity and deviation increases with $\alpha$. But the deviation increases with a slightly greater rate than the fidelity. Therefore, the $F/D$ ratio goes down with increasing $\alpha$.  Furthermore, $F>2/3$ when $\alpha \gtrsim 0.05$, so we plot Fig. \ref{fig:bitflip} for the same.

\subsubsection{Phase-flip noise}
\label{sec:phaseflip}
The phase-flip noise is modeled by the action of the Pauli operator $\sigma^z$. 
It keeps the state $\ket{0}$ as it is and adds a phase of $e^{i\pi} = -1$ to $\ket{1}$. Given an arbitrary state, $\rho$, it keeps it unaltered with a probability $p$, and phase-flips it
with a probability of $1 - p$. Again, $0 \leq p \leq 1$. Note that the Krauss operators for phase-flip noise can be written as
\begin{eqnarray}
K_0^{\text{pf}} &=& \sqrt{pq} ~\mathbb{I}\otimes \mathbb{I}, ~K_1^{\text{pf}} = \sqrt{p(1-q)} ~\mathbb{I} \otimes \sigma^z, \nonumber \\
K_2^{\text{pf}} &=& \sqrt{(1-p)q} ~\sigma^z  \otimes \mathbb{I}, ~K_3^{\text{pf}} = \sqrt{(1-p)(1-q)} ~\sigma^z \otimes \sigma^z, \nonumber \\
\end{eqnarray}
where the superscript pf naturally denotes the phase-flip noise and $\sigma_{1(2)}^z = \begin{bmatrix}
1 & 0 \\
0 & -1
\end{bmatrix}$. The corresponding state evolves as
\begin{eqnarray}
\rho \xrightarrow{\text{phase-flip}} \sum_{i=0}^3 K_i^{\text{pf}} \rho (K_i^\text{pf})^\dagger.
\end{eqnarray}

Therefore in the presence of phase-flip noise in both the qubits with probabilities $p$ and $q$, an arbitrary resource state $\rho_{12}$ becomes  
\begin{eqnarray}
\rho \xrightarrow{\text{phase flip}} pq \rho_{12} &+& (1-p)q \sigma^z_1 \rho_{12} \sigma_1^z + p(1-q) \sigma^z_2 \rho_{12} \sigma_2^z \nonumber \\ &+& (1-p)(1-q) \sigma^z_1 \sigma^z_2 \rho_{12} \sigma^z_2 \sigma_1^z .
\end{eqnarray} 
When an arbitrary state, $|\eta\rangle =$ $\cos \frac{\theta}{2}|0\rangle +e^{i\phi} \sin \frac{\theta}{2}|1\rangle$, is to be teleported through such a channel, the fidelity of the output state with Bob with $|\eta\rangle$ reads
\begin{eqnarray}
F^\eta &=& \big(pq +(1-p)(1-q)\big) \big[ 1 - \frac{1}{2}(1 - 2 \sqrt{\alpha(1-\alpha)})\sin^2 \theta\big] \nonumber \\
&+& \frac{1}{2}(p + q - 2pq)\big[ 1 - \frac{1}{2}(1 + 2 \sqrt{\alpha(1-\alpha)})\sin^2 \theta\big].
\end{eqnarray}

Again for simplicity, we assume $q = 1$. The expressions of fidelity and fidelity deviation reads
\begin{widetext}
\begin{eqnarray}
F^{\text{phase-flip}} &=& \frac{2}{3} \big( 1+ (2p-1)  \sqrt{(1-\alpha ) \alpha} \big) \nonumber \\
D^{\text{phase-flip}} &=& \frac{\sqrt{1+4 (2 p-1)^2 \alpha -4 (2 p-1)^2 \alpha ^2-4 \sqrt{(1-\alpha ) \alpha }+8 (1-p) \sqrt{(1-\alpha ) \alpha }}}{3 \sqrt{5}} \nonumber \\
 &=& \frac{1-2(2p-1)\sqrt{(1-\alpha ) \alpha }}{3\sqrt{5}} = \frac{1-F^{\text{phase-flip}}}{\sqrt{5}}
\label{eq:phaseflip_F,D}
\end{eqnarray}
\end{widetext}
It is clear from the above expressions that $D^{\text{phase-flip}}$ decreases with increasing $F^{\text{phase-flip}}$. Therefore, like the noiseless setting, the usual ordering of resource states is retained in the presence of phase-flip noise.

\subsubsection{Bitphase-flip noise}
\label{sec:bitphaseflip}
The bitphase-flip noise is modeled by the action of the Pauli operator $\sigma^y$. We consider an arbitrary resource state $\rho_{12}$ suffers bitphase-flip noise in both the qubits with probabilities $p$ and $q$. Note that the Krauss operators for bitphase-flip noise can be written as
\begin{eqnarray}
K_0^{\text{bpf}} &=& \sqrt{pq} ~\mathbb{I}\otimes \mathbb{I}, ~K_1^{\text{bpf}} = \sqrt{p(1-q)} ~\mathbb{I} \otimes \sigma^y, \nonumber \\
K_2^{\text{bpf}} &=& \sqrt{(1-p)q} ~\sigma^y  \otimes \mathbb{I}, ~K_3^{\text{bpf}} = \sqrt{(1-p)(1-q)} ~\sigma^y \otimes \sigma^y, \nonumber \\
\end{eqnarray}
where the superscript ``bpf" naturally denotes the bitphase-flip noise and $\sigma_{1(2)}^y = \begin{bmatrix}
0 & -i \\
i & 0
\end{bmatrix}$. The corresponding state evolves as
\begin{eqnarray}
\rho_{12} \xrightarrow{\text{bitphase-flip}} \sum_{i=0}^3 K_i^{\text{bpf}} \rho_{12} (K_i^\text{bpf})^\dagger.
\end{eqnarray}
Now, $\rho_{12}$ evolves
\begin{eqnarray}
\rho_{12} \xrightarrow{\text{bitphase-flip}} pq \rho_{12} &+& (1-p)q \sigma^y_1 \rho_{12} \sigma_1^y + p(1-q) \sigma^y_2 \rho_{12} \sigma_2^y \nonumber \\ &+& (1-p)(1-q) \sigma^y_1 \sigma^y_2 \rho_{12} \sigma^y_2 \sigma_1^y .
\end{eqnarray} 
Note that the expressions of fidelity for an arbitrary input state to be teleported is same as that in for the bit-flip noise (see Eq. \ref{eq:fid_bitflip}). Consequently the average fidelity and its deviation are also exactly same. Therefore, the physics of teleportability score, remains exactly the same as in case of bit-flip noise (see Fig. \ref{fig:bitflip}).

\begin{figure*}
\includegraphics[width=0.95\linewidth]{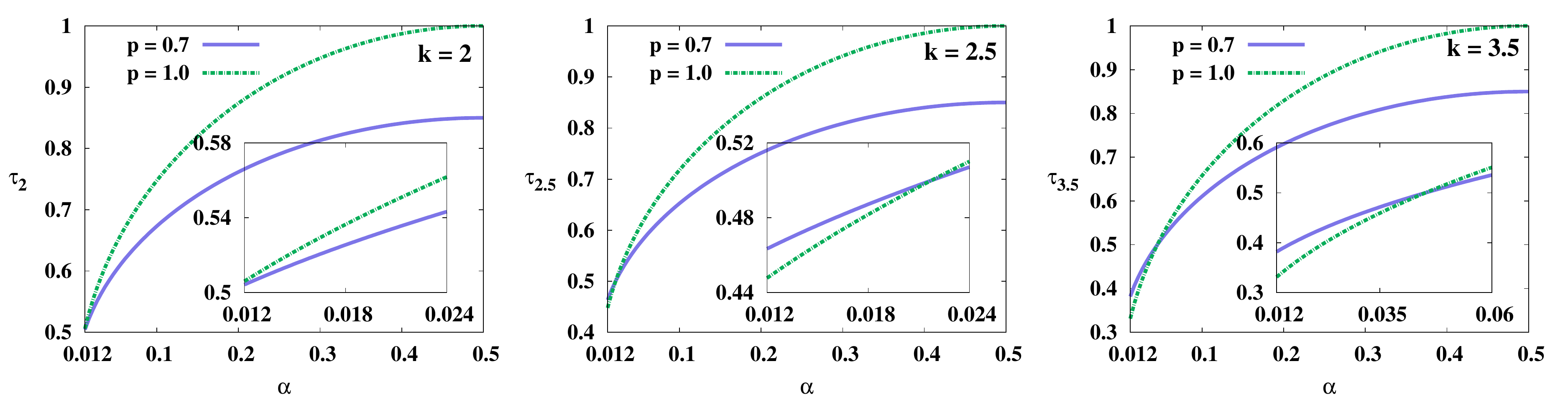}
\caption{Teleportability score when resource states suffer from global depolarizing noise as compared to the noiseless value for $k=2, 2.5$, and $3.5$. For $k = 2.5$ and $3.5$, we get a finite range of  $\alpha$ where the noisy states outperform the noiseless one. This range is enhanced for larger $k$ values. The $k = 2$ is the marginal case above which we observe this feature. Note that the $\alpha$ offset of $0.012$ ensures that we sample only those states which yield nonclassical teleportation for the given noise settings of $p = 0.7$, and $q = 1$.
All axes are dimensionless.}
\label{fig:global1}
\end{figure*}

\subsection{Global noise}
\label{sec:global}
Apart from the local noise models considered in the previous section, the resource state might also suffer from global noise. In such a situation, the environment interacts with the whole composite system. In this section, we analyze the response of teleportability score to global depolarizing noise where the whole bipartite state becomes a convex mixture of itself with white noise. We also consider the combined effect of both local and global depolarizing noise on the teleportability score of the resource state.

\subsubsection{Global depolarizing noise}
A bipartite entangled state $|\psi ^{\alpha}\rangle = \sqrt{\alpha} |00\rangle + \sqrt{1-\alpha} |11\rangle$ under the action of a global depolarizing channel gets admixed with a maximally mixed state, and becomes
\begin{equation}
\rho ^{\alpha,p}= p \rho ^{\alpha}+ (1-p) \frac{\mathbb{I}\otimes \mathbb{I}}{4},
\end{equation}
where $\rho ^{\alpha}= |\psi ^{\alpha}\rangle \langle \psi^{\alpha} |$ and $p$ is the noise parameter with $0 \leq p \leq 1$. When an arbitrary state, $|\eta\rangle =$ $\cos \frac{\theta}{2}|0\rangle +e^{i\phi} \sin \frac{\theta}{2}|1\rangle$, is to be teleported through such a channel, the fidelity of the output state with Bob with $|\eta\rangle$ reads 
\begin{eqnarray}
F^\eta = p\big[ 1 - \frac{1}{2}(1 - 2 \sqrt{\alpha(1-\alpha)})\sin^2 \theta\big] + \frac{1-p}{2}.
\end{eqnarray}

The average teleportation fidelity $F^{\text{dep}} (p)$ can estimated as
\begin{eqnarray}
F^{\text{dep}}(p)&=&\frac{2p}{3}\left(1+ \sqrt{\alpha (1-\alpha)}\right)+ \frac{1-p}{2} \nonumber \\
&=& pF + \frac{1-p}{2},
\end{eqnarray} 
where $F = \frac{2}{3}+\frac{2 \sqrt{\alpha(1-\alpha)}}{3}$, the average fidelity in the noiseless case, see Eq. \ref{eq:fid_noiseless}. Similarly, the fidelity deviation reads
\begin{eqnarray}
D^{\text{dep}} (p) &=& \frac{p\sqrt{1-4\sqrt{\alpha (1-\alpha)} + 4\alpha (1-\alpha)}}{3\sqrt{5}} \nonumber \\
&=& \frac{p(1-2\sqrt{\alpha (1-\alpha)})}{3\sqrt{5}} = \frac{p(1-F)}{\sqrt{5}}.
\end{eqnarray}
Note that the deviation in fidelity is lower compared to the noiseless case. This allows for the possibility that for high sensitivity requirements, the teleportability score in the presence of global depolarizing noise becomes better compared to the noiseless case as shown in Fig. \ref{fig:global1} for a typical example with $p =0.7$. Firstly note that the we need an offset in the $\alpha$ value, say $\alpha_{cl}$ which guarantees non-classical teleportation, which in this case turns out to be, $\alpha_{cl} = 0.012$. Next, let us denote $\alpha_n^k$ to be the $\alpha$ value upto which the noisy case is ``better” than the noiseless case for a fixed $k$. Therefore, $\alpha_c$ to $\alpha_n^k$ is the region of interest. For $p = 0.7$, the $k = 2$ gives the marginal case where $\alpha_n^{k = 2} = \alpha_{cl}$. For $k>2$ we enter the regime where one does get a finite  range of $\alpha$ where noise state turns out to be ``better". We now tabulate $\alpha_n^k$ for some typical $k$ values in Table. \ref{tab:tab1}.
As expected, the $\alpha_n^k$ values, and correspondingly the $\alpha$-range which gives better teleportation for the noisy state grows with $k$. Lastly, the $k^*$ value in this case is around $8$, so the $k$-values chosen for the analysis are well within the bounds. 

\begin{table}[ht]
\begin{tabular}{|c|c|}
\hline
$k$ & $\alpha_n^k$ \\ \hline
$2.1$ & $0.013 ~(\alpha_{cl} + 0.001)$ \\ \hline
$2.5$ & $0.022 ~(\alpha_{cl} + 0.010)$ \\ \hline
$3.5$ & $0.033 ~(\alpha_{cl} + 0.021)$ \\ \hline
$4.0$ & $0.056 ~(\alpha_{cl} + 0.044)$ \\ \hline
\end{tabular}
\caption{Enhancement of $\alpha_n^k$ values with $k (>2)$ for depolarizing noise with $p=0.7$. For a graphical representation of the $k = 2.5$ and $3.5$ cases, see Fig. \ref{fig:global1}. Note that in this case, $\alpha_{cl} = 0.012$.}
\label{tab:tab1}
\end{table} 
 
\subsubsection{Joint action of local and global depolarizing noise}
Here we consider a scenario where both qubits are both locally as well as globally affected by depolarizing noise. Naturally, there will three independent noise parameters $p$ (for global noise), $p_{1}$ and $p_{2}$ (for local noises). The initial resource state ($\rho^\alpha = |\psi^\alpha\rangle$) under the action of this channel would become
\begin{eqnarray}
\rho ^{\alpha, p_{1},p_{2},p}&=& p\frac{\mathbb{I}\otimes \mathbb{I}}{4}+ \frac{p_{1}}{2}\left\lbrace \mathbb{I}\otimes \text{Tr}_{1}(\rho ^{\alpha})\right \rbrace + \frac{p_{2}}{2}\left \lbrace \text{Tr}_{2}(\rho ^{\alpha})\otimes \mathbb{I}\right \rbrace \nonumber \\
&+& (1-p_{1}-p_{2}-p)\rho ^{\alpha}
\end{eqnarray}
The average fidelity $F^{\text{dep-local-global}}(p,p_{1},p_{2})$ can be estimated as 
\begin{eqnarray}
F^{\text{dep-local-global}}&(&p,p_{1},p_{2})=\frac{2}{3}\left(1+(1-p)\sqrt{\alpha (1-\alpha )}\right) \nonumber \\
&-& \frac{(p_{1}+p_{2})\left( 1+4\sqrt{\alpha (1-\alpha )}\right)}{8}-\frac{p}{6}
\end{eqnarray}
The corresponding fidelity deviation $D^{\text{dep-local-global}}(p,p_{1},p_{2})$ reads 
\begin{widetext}
\begin{eqnarray}
D^{\text{dep-local-global}}(p,p_{1},p_{2}) &=& \frac{1}{24\sqrt{5}}\left[ \left(64-96 p_{2} + 51 p_{2}^{2}\right)+4\left(64 + 3 p_{2} (7 p_{2} -32)\right)\alpha (1-\alpha) - (256-384 p_{2} + 144 p_{2}^{2})\sqrt{\alpha (1-\alpha)} \right. \nonumber \\
&+&\left\lbrace 64 p^{2} + 32p (3 p_{1}+ 3 p_{2} -4) - 96 p_{1}\right\rbrace \left\lbrace 1- 4\sqrt{\alpha (1-\alpha)} + 4\alpha (1-\alpha)\right\rbrace \nonumber \\
 &+&\left. p_{1}^{2} \left\lbrace 51 - 12\sqrt{\alpha} (-7\sqrt{\alpha} +7\alpha +12 \sqrt{1-\alpha} ) \right\rbrace +p_{2} \left\lbrace -17 + 4\sqrt{\alpha} (-7\sqrt{\alpha} +7 \alpha +12 \sqrt{1-\alpha})\right\rbrace \right]^{\frac{1}{2}}.
\end{eqnarray}
\end{widetext}

In summary, for the class of states considered in this work, the teleportation fidelity grows monotonically with the entanglement content of the state. When these states are subjected to local noise, the same monotonicity of fidelity with respect to the entanglement content of the parent state is observed. This is the usual ``ordering of states," where a higher entangled initial state always (in presence and absence of noise) leads to better teleportation if only analyzed via average fidelity. However, what we show is that such ordering is not always valid when one incorporates fidelity deviation into the picture and quantifies the performance of teleportation by the teleportability score. Higher entangled states can possess lower values of teleportability score values owing to the large increase of fluctuations in the presence of local noise. Therefore the usual ``ordering" of states with respect to entanglement does not remain valid anymore.

\section{Conclusion}
\label{sec:conclusion}
Traditionally, the performance of teleportation is calibrated by the average fidelity it yields. However, not all states are teleported equally, and the fidelity of each state can be widely dispersed for a fixed value of average fidelity. These fluctuations turn out to be detrimental during various quantum information processing tasks like implementation of quantum gates etc. Therefore, the characterization of the performance of teleportation via the average fidelity alone is very restrictive. In this work, we characterize teleportation via both average fidelity ($F$) and fidelity deviation ($D$), and reanalyze the usual teleportation protocol in presence of local and global noises.

In this work, we define teleportability score, $\tau$, as the difference between average fidelity and the pay off term due to fluctuations, $F-k.D$, where, $k$ denotes the sensitivity of the score to fluctuations in fidelity. In the noiseless scenario, we find that for any value of $k$, $\tau_k$ is a monotonically increasing function of fidelity for the  pure resource states considered in our analysis. In the presence of local bit-flip and bitphase-flip noise, if high sensitivity to fluctuations is imposed, although the fidelity alone increases with increasing entanglement of the shared resource state, the teleportability score on the other hand might decrease. It therefore reverses the known hierarchy of resource states in terms of their capability of teleportation. When the resource state is affected by global depolarizing noise, for low values of entanglement and high sensitivity demands, the noisy states can sometime outperform the noiseless ones in terms of the teleportability score.

Note, although we have tried to provide an upper bound to the sensitivity requirements, we want to highlight that we think it is very difficult to put a bound on $k$ from purely a rigorous theoretical perspective. Let us illustrate this difficulty via an example. Suppose one has to build a highway. Cars travel on the highway with high speed so there is a possibility that they experience some bump or fluctuations and fall off the side end of the highway causing an accident. Such accidents can be minimized if the highway is wide enough. But now if one asks the question: how wide is wide enough for the highway, then one runs into trouble since, just like in our teleportation case, it is very difficult to provide a rigorous mathematical criterion for the width. In such a situation one has to resort to some practical reasoning to come up with a safe width, for example by considering the traffic rate, number of big vehicles per day, cost of increasing the width, availability of free land etc. This practicality strategy is what we employ in this case also to compute a reasonable upper bound for $k$.

We also want to mention a possible utility of our teleportation characterizing scheme from an experimental point of view. We illustrate this via a simple example. Consider a quantum circuit where teleportation is used as an intermediate step to transfer unknown quantum states amongst nodes of the circuit. Suppose that in the receiving node, the quantum information (the unknown quantum state) has to be processed by a quantum gate. Now as pointed out in \cite{gatefid2}, the performance of the quantum gates are sensitive to fluctuations in its input, which in this case is the state it receives after teleportation. Typically the resource state that is used for teleportation is non-maximally entangled and it inevitably suffers from environmental noise. So, on average, the performance of the quantum gate would greatly suffer fluctuations in fidelity of the state coming out of the teleportation process. Therefore, under the constraint that one uses states of a given (non maximal) amount of entanglement, which suffer from noise, calibrating the performance of this teleportation protocol via both fidelity and fidelity deviation becomes essential. Furthermore, note that all quantum gates will not be equally sensitive to fluctuations in fidelity of its input. This motivates the choice of a measure like teleportability score to rate the performance of quantum teleportation which would help to capture the information about the fluctuations and thereby aid in providing a better characterization of the performance of the quantum circuit in general.

Although,  we have analyzed the performance of teleportation by considering both fidelity and its deviation, similar analysis can also be carried out for other protocols whose performance is  quantified by the average fidelity.

\section*{Acknowledgement}
The authors thank Debarshi Das, Aditi Se (De), Ujjwal Sen, and  Somshubhro Bandyopadhyay for carefully going through the manuscript and sharing invaluable comments.

%
%
%
%


\begin{thebibliography}{99}


\bibitem{BB84} C. H. Bennett  and G. Brassard, Proceedings of IEEE International Conference on Computers, Systems and Signal Processing, Bangalore, India (IEEE, NY, 1984).


\bibitem{Crypto} A. Ekert, Phys. Rev. Lett. {\bf 67}, 661 (1991); N. Gisin, G. Ribordy, W. Tittel, and H. Zbinden, Rev. Mod. Phys. {\bf 74}, 145 (2002); 
V. Scarani, H. Bechmann-Pasquinucci, N. J. Cerf, M. Du\v sek, N. L\" utkenhaus, and M. Peev,
Rev. Mod. Phys. {\bf 81}, 1301 (2009).

\bibitem{DC} C. H. Bennett, and S. J. Wiesner, Phys. Rev. Lett. {\bf 69}, 2881 (1992); S. Bose, M. B. Plenio, and V. Vedral, J. Mod. Opt. {\bf 47}, 291 (2000); T. Hiroshima, J. Phys. A: Math. Gen. {\bf 34}, 6907 (2001); G. Bowen, Phys. Rev. A {\bf 63}, 022302 (2001); M. Horodecki, P. Horodecki, R. Horodecki, D. Leung, and B. Terhal, Quantum Inf. and Comput. {\bf 1}, 70 (2001); X. S. Liu, G. L. Long, D.M. Tong, and F. Li, Phys. Rev. A {\bf 65}, 022304 (2002); M. Ziman and V. Bu\^zek, ibid. {\bf 67}, 042321 (2003); D. Bru{\ss}, G. M. D'Ariano, M. Lewenstein, C. Macchiavello, A. Sen(De), and U. Sen,
Phys. Rev. Lett. \textbf{93}, 210501 (2004); For a recent review on quantum communication, see e.g. A. Sen(De) and U. Sen, Physics News {\bf 40}, 17 (2010) (arXiv:1105.2412).

\bibitem{error_correction} A. R. Calderbank and P. W. Shor, Phys. Rev. A {\bf 54}, 1098 (1996); A. M. Steane, Phys. Rev. Lett. {\bf 77}, 793 (1996); A. M. Steane, Phys. Rev. A {\bf 54}, 4741 (1996); A. R. Calderbank, E. M. Rains, P. W. Shor, and N. J. A. Sloane, Phys. Rev. Lett. {\bf 78}, 405 (1997).

 \bibitem{bennett1993}
C. H. Bennett, G, Brassard, C. Cr\'epeau, R. Jozsa, A. Peres,
 and W. K. Wootters,
 Phys. Rev. Lett. \textbf{70,}  1895 (1993).
 

\bibitem{communication_exp} C. H. Bennett,  F. Bessette, G. Brassard, L. Salvail, and J. Smolin,  J. Cryptology {\bf 5}, 3 (1992); K. Mattle, H. Weinfurter, P. G. Kwiat and A. Zeilinger, Phys. Rev. Lett. {\bf 76}, 4656 (1996); D. Leibfried, R. Blatt, C. Monroe, and D. Wineland, Rev. Mod. Phys. {\bf 75}, 281 (2003); L. M. K. Vandersypen and I. L Chuang, Rev. Mod. Phys. {\bf 76}, 1037 (2005); 
H. Hafner, C. F. Roose, and R. Blatt, Phys. Rep. {\bf 469}, 155 (2008); 
K. Singer, U. Poschinger, M. Murphy, P. Ivanov, F. Ziesel, T. Calarco, and F. Schmidt-Kaler, Rev. Mod. Phys. {\bf 82}, 2609 (2010); L.-M. Duan, and C. Monroe, Rev. Mod. Phys. {\bf 82}, 1209 (2010); 
 and references
therein.



 
  \bibitem{tele_review}
J.-W. Pan, Z.-B. Chen, C.-Y. Lu, H. Weinfurter, A. Zeilinger, and M \.Zukowski, Rev. Mod. Phys. {\bf 84}, 777 (2012); S. Pirandola, J. Eisert, C. Weedbrook, A. Furusawa, and S. L. Braunstein, Nat. Photonics {\bf 9}, 641 (2015).

\bibitem{murao1998}
 M. Murao, M. Plenio, S. Popescu, V. Vedral, and P. L. Knight, Phys. Rev. A \textbf{57}, 4075, (1998); M. Murao, D. Jonathan, M. B. Plenio, and V. Vedral, Phys. Rev. A \textbf{59}, 156 (1999);
 Z. Zhao, A.-N. Zhang, X.-Q. Zhou, Y.-A. Chen, C.-Y. Lu, A. Karlsson, and J.-W. Pan, Phys. Rev. Lett. \textbf{95}, 030502 (2005);  S. Koike, H. Takahashi, H. Yonezawa, N. Takei, S.  L. Braunstein, T. Aoki, and A. Furusawa, Phys. Rev. Lett. \textbf{96}, 060504 (2006); M. Radmark, M. \.Zukowski, and M. Bourennane, New J. Phys. \textbf{11}, 103016 (2009); A. Sen(De) and U. Sen, Phys. Rev. A \textbf{81}, 012308 (2010); J. Lee, S.-W. Ji, J. Park, and H. Nha, Phys. Rev. A {\bf 94}, 062318 (2016); M. M. Cunha, E. A. Fonseca, and F. Parisio, \emph{Non-ideal teleportation of tripartite entanglement: Einstein-Podolsky-Rosen versus Greenberger-Horne-Zeilinger schemes}, arXiv:1611:01167.

   

%

%
%



%
%
%
%
%
%


 



\bibitem{tele_photon_exp} D. Bouwmeester, J.-W. Pan, K. Mattle, M. Eibl, H. Weinfurter and A. Zeilinger,
Nature {\bf 390}, 575 (1997); D. Boschi, S. Branca, F. De Martini, L.  Hardy, and S. Popescu,
  Phys. Rev. Lett. \textbf{80,}  1121 (1998); X.-S. Ma, T. Herbst, T. Scheidl, D. Wang, S. Kropatschek, W. Naylor, B. Wittmann, A. Mech, J. Kofler, E. Anisimova, V. Makarov, T. Jennewein, R. Ursin and A. Zeilinger,
Nature {\bf 489}, 269 (2012).

  \bibitem{tele_photon_exp2} Y.-H. Kim, S. P. Kulik and Y. Shih, Phys. Rev. Lett. {\bf 86}, 1370 (2001); J. Yin, J.-G. Ren, H. Lu, Y. Cao, H.-L. Yong, Y.-P. Wu, C. Liu, S.-K. Liao, F. Zhou, Y. Jiang, X.-D. Cai, P. Xu, G.-S. Pan, J.-J. Jia, Y.-M. Huang, H. Yin, J.-Y. Wang, Y.-A. Chen, C.-Z. Peng and J.-W. Pan,
  Nature {\bf 488}, 185 (2012).

\bibitem{tele_contvar_exp} A. Furusawa, J. L. S\o rensen, S. L. Braunstein, C. A. Fuchs, H. J. Kimble, E. S. Polzik, Science {\bf 282}, 706 (1998); W. P. Bowen, et al., Phys. Rev. A {\bf 67}, 032302 (2003);  T. C. Zhang, K. W. Goh, C. W. Chou, P. Lodahl, and H. J.  Kimble, Phys. Rev. A {\bf 67}, 033802 (2003); N. Takei,  H. Yonezawa, T. Aoki, and A. Furusawa, Phys. Rev. Lett. {\bf 94}, 220502 (2005); H. Yonezawa, S. L. Braunstein, and A. Furusawa, Phys. Rev. Lett. {\bf 99}, 110503 (2007); N. Takei, et al., Phys. Rev. A {\bf 72}, 042304 (2005); N. Lee, et al., Science {\bf 332}, 330 (2011); M. Yukawa, H. Benichi, and A. Furusawa,  Phys. Rev. A {\bf 77}, 022314 (2008).

%

  
  \bibitem{tele_ion_exp} M. D. Barrett, J. Chiaverini, T. Schaetz, J. Britton, W. M. Itano, J. D. Jost, E. Knill, C. Langer, D. Leibfried, R. Ozeri and D. J. Wineland,  Nature {\bf 429}, 737 (2004); M. Riebe, H. H\" affner, C. F. Roos, W. H\" ansel, J. Benhelm, G. P. T. Lancaster, T. W. K\" orber, C. Becher, F. Schmidt-Kaler, D. F. V. James and R. Blatt, Nature {\bf 429}, 734 (2004); S. Olmschenk, D. N. Matsukevich, P. Maunz, D. Hayes, L. M. Duan, and C. Monroe,  Science \textbf{323,} 486 (2009).
  
   \bibitem{tele_exp_coldatom}
 N. Solmeyer, X. Li, and Q. Quraishi, Phys. Rev. A
  \textbf{93}, 042301 (2016).
  
  
  \bibitem{tele_nmr_exp} M. A. Nielsen, E. Knill and R. Laflamme,  Nature {\bf 396}, 52 (1998).
  
  \bibitem{tele_supercond} L. Steffen, Y. Salathe, M. Oppliger, P. Kurpiers, M. Baur, C. Lang, C. Eichler, G. Puebla-Hellmann, A. Fedorov and A. Wallraff, Nature {\bf 500}, 319 (2013).
  




%
%
%
%
%
%
%
%
%

\bibitem{ref11} H. Jeong and M. S. Kim
Phys. Rev. A {\bf 65}, 042305 (2002).

\bibitem{ref12} T. C. Ralph, A. Gilchrist, G. J. Milburn, W. J. Munro, and S. Glancy, Phys. Rev. A {\bf 68} 042319 (2003).

\bibitem{gatefid1} Pedersen L H, M{\o}ller N M and M{\o}lmer K, Phys. Lett. A {\bf 372}, 7028 (2008).

\bibitem{gatefid2} E. Magesan, R. Blume-Kohout, and J. Emerson, Phys. Rev. A {\bf 84}, 012309 (2011).
 
 \bibitem{dev1} J. Bang, S. W. Lee, H. Jeong, and  J. Lee,  Phys. Rev. A {\bf 86}, 062317 (2012).
  
  
 \bibitem{fid_dev} J. Bang, J. Ryu, and D. Kaszlikowski, J. Phys. A: Math. Theor. {\bf 51}, 135302 (2018).
 
\bibitem{g1} A. Ghosal, D. Das, S. Roy, and S. Bandyopadhay, J. Phys. A: Math. Theor. {\bf 53}, 145304 (2020). 

\bibitem{g2} A. Ghosal, D. Das, S. Roy, and S. Bandyopadhyay, 
Phys. Rev. A {\bf 101}, 012304 (2020).
 
 \bibitem{preskil} J. Preskill, Lecture Notes, available at http://www.theory.caltech.edu/people/preskill/ph219/.

\bibitem{Nielsen_chuang}  M. A. Nielsen and I. L. Chuang, {\it Quantum Computation and Quantum Information} (Cambridge University Press, Cambridge, 2000).
 
  
 \bibitem{horodecki2009}
  R. Horodecki, P. Horodecki, M. Horodecki, and K. Horodecki, Rev. Mod. Phys. \textbf{81}, 865 (2009).
  
  \bibitem{classical1}
  S. Massar and S. Popescu, Phys. Rev. Lett. \textbf{74}, 1259 (1995).
  
  \bibitem{classical2}
  N. Gisin, Phys. Lett. A {\bf 210}, 3 (1996).

\bibitem{Schmidt_decomposition} A. Peres, {\it Quantum Theory: Concepts and Methods}  (Kluwer,
Dordrecht, 1993 , pp. 131-133).

  
  
\bibitem{horo_fid} P. Horodecki, M. Horodecki, R. Horodecki,  	Phys. Rev. A \textbf{60}, 1888 (1998).

\bibitem{Bell-basis} The Bell basis consists of four maximally entangled states, given by
 \(|\phi^{\pm}\rangle = \frac{1}{\sqrt{2}}(|00\rangle \pm |11\rangle)\),
 and
 \(|\psi^{\pm}\rangle = \frac{1}{\sqrt{2}}(|01\rangle \pm |10\rangle)\),
 where $\{\ket{0},\ket{1}\}$ is the computational basis.

 \bibitem{q_repeater}
  H. Briegel, W. Dür, J. I. Cirac, and P. Zoller, Phys. Rev. Lett. \textbf{81,} 5932 (1998);  J. Dias, and T. C. Ralph, Phys. Rev. A \textbf{95,} 022312 (2017).

  
   
 
%
%
%
%
%
%
%
%
%
%
%
%
%
%
%
%
%
%
%
%
%
%
%


  \end{thebibliography}
\end{document}